\documentclass[12pt]{article}

\usepackage{graphicx}

\usepackage{cite}

\begin{document}

\title{On Dalgarno and Lewis Perturbation Theory for Scattering States}
\author{Paolo Amore \\
Facultad de Ciencias, Universidad de Colima,\\
Bernal D\'{i}az del Castillo 340, Colima, Colima,\\
Mexico.
 \and Francisco M. Fern\'{a}ndez \\
INIFTA (Conicet,UNLP), Divisi\'on Qu\'imica Te\'orica, \\
Diag. 113 y 64 S/N, Sucursal 4, Casilla de Correo 16, \\
1900 La lata, Argentina}

\maketitle

\begin{abstract}
We apply the method of Dalgarno and Lewis to scattering states and discuss
the choice of the unperturbed model in order to have a convergent
perturbation series for the phase shift.
\end{abstract}

\section{Introduction \label{sec:intro}}

There has been some interest in the application of on--shell perturbation
theory to simple scattering problems in one dimension \cite
{ACCLY92,NBPT96,ACC97}. Those approaches were mainly based on logarithmic
perturbation theory\cite{ACCLY92} and the Dalgarno--Lewis perturbation theory%
\cite{NBPT96,ACC97}. In order to bypass the problem posed by the nodes of
real wavefunctions they used complex wavefunctions (Jost functions).
Although the method of Dalgarno and Lewis was proposed as a device to
calculate sum over states\cite{DL55} it proves to be an efficient approach
for on--shell perturbation theory\cite{BO78,F01}. This method has recently
been applied to solve the differential equation for the density distribution
in scattering problems\cite{MW01}.

The purpose of this paper is to discuss the application of the Dalgarno and
Lewis perturbation theory to potential scattering using real wavefunctions.
We also consider the choice of the unperturbed or reference model in order
to have convergent perturbation series for the phase shift. In Sec.~\ref
{sec:model} we introduce the model. In Sec.~\ref{sec:method} we propose an
alternative application of the method of Dalgarno and Steward and a choice
of the unperturbed model potential. In Sec.~\ref{sec:example} we apply the
main equations to an illustrative example, and in Sec.~\ref{sec:conclusions}
we summarize the main results of the paper and draw some conclusions.

\section{The model \label{sec:model}}

For simplicity we consider the one--dimensional model
\begin{equation}
-\frac{\hbar ^{2}}{2m}\psi ^{\prime \prime }(r)+V(r)\psi (r)=E\psi (r)
\label{eq:Schro}
\end{equation}
and choose the boundary condition $\psi (0)=0$ in order to mimic the
s--states of a central--field model. The prime denotes differentiation with
respect to the coordinate. The potential--energy function $V(r)$ is negative
for $r<L$ and zero for $r\geq L$. In order to derive a dimensionless
differential equation we define the new independent variable $x=r/L$ and
rewrite Eq. (\ref{eq:Schro}) as
\begin{equation}
y^{\prime \prime }(x)=[v(x)-\epsilon ]y(x)  \label{eq:y"}
\end{equation}
where
\begin{equation}
v(x)=\frac{2m}{\hbar ^{2}}V(Lx),\;\epsilon =\frac{2m}{\hbar ^{2}}%
E,\;y(x)=\psi (Lx).  \label{eq.v(x),e}
\end{equation}
The boundary conditions are
\begin{equation}
y(0)=0,\;y(x)=\frac{1}{k}\sin (kx+\delta ),\;x>1  \label{eq:bc_y}
\end{equation}
where $\delta $ is the phase shift and $k=\sqrt{\epsilon }$. In order to
avoid ambiguities regarding to multiples of $\pi $ we choose $\delta (k)$ to
satisfiy $\delta (\infty )=0$.

In order to apply perturbation theory we choose a closely related problem
\begin{equation}
y_{0}^{\prime \prime }(x)=[v_{0}(x)-\epsilon ]y_{0}(x)  \label{eq:y0"}
\end{equation}
with similar boundary conditions
\begin{equation}
y_{0}(0)=0,\;y_{0}(x)=\frac{1}{k}\sin (kx+\delta _{0}),\;x>1.
\label{eq:bc_y0}
\end{equation}
As it is custommary in most applications of perturbation theory we write $%
v(x)=v_{0}(x)+\lambda v_{1}(x)$, where the perturbation parameter $\lambda $
is set equal to unity at the end of the calculation, and expand
\begin{eqnarray}
y(x) &=&\sum_{j=0}^{\infty }y_{j}(x)\lambda ^{j},  \nonumber \\
\delta &=&\sum_{j=0}^{\infty }\delta _{j}\lambda ^{j}.  \label{eq:PT_series}
\end{eqnarray}

\section{The method of Dalgarno and Lewis \label{sec:method}}

For simplicity we write $Q(x)=v(x)-\epsilon $, and $Q_{0}(x)=v_{0}(x)-%
\epsilon $. The function $F(x)=y(x)/y_{0}(x)$ satisfies the differential
equation
\begin{equation}
\left( y_{0}(x)^{2}F^{\prime }(x)\right) ^{\prime }=\Delta
Q(x)F(x)y_{0}(x)^{2}  \label{eq:diff_eq_F}
\end{equation}
where $\Delta Q=Q-Q_{0}$. Since the nodes of $y(x)$ and $y_{0}(x)$ do not
coincide one expects $F(x)$ to have poles at the nodes of $y_{0}(x)$.
However, $F(x)$ does not appear in the resulting expression for $y(x)$ (see
below) and, consequently, that problem does not arise in practical
applications of the method of Dalgarno and Lewis.

Notice that
\begin{equation}
y_{0}(x)^{2}F^{\prime }(x)=y^{\prime }(x)y_{0}(x)-y(x)y_{0}^{\prime
}(x)=W(y,y_{0})(x)  \label{eq:Wronsk1}
\end{equation}
is the Wronskian of the perturbed and unperturbed solutions. On integrating
Eq. (\ref{eq:diff_eq_F}) we have
\begin{equation}
W(y,y_{0})(x)=\int_{0}^{x}\left( \Delta Qyy_{0}\right) (x^{\prime
})\,dx^{\prime }.  \label{eq:Wronsk2}
\end{equation}
At $x=1$ we substitute the asymptotic forms of $y(x)$ and $y_{0}(x)$ into
the Wronskian and obtain an expression for the phase shift
\begin{equation}
\sin (\Delta \delta )=-k\int_{0}^{1}\left( \Delta Qyy_{0}\right) (x)\,dx.
\label{eq:sin_phase}
\end{equation}

If we integrate Eq. (\ref{eq:Wronsk2}) between $x=1$ and $x$ and multiply
the result by $y_{0}(x)$ we obtain an expression for $y(x)$:
\begin{equation}
y(x)=Cy_{0}(x)+y_{0}(x)\int_{1}^{x}\frac{dx^{\prime }}{y_{0}(x^{\prime })^{2}%
}\int_{0}^{x^{\prime }}\left( \Delta Qyy_{0}\right) (x^{\prime \prime
})\,dx^{\prime \prime }  \label{eq:y(x)}
\end{equation}
where $C$ is an integration constant. Notice that this expression is free
from poles because the function $F(x)$ does not appear explicitly in it.

When $x>1$ then $x^{\prime }>1$ and we substitute Eq. (\ref{eq:sin_phase})
for the second integral in Eq. (\ref{eq:y(x)}). Moreover, in order to have
the correct asymptotic expression for $y(x)$ the integration constant should
be
\begin{equation}
C=\sin (\Delta \delta )\cot (k+\delta _{0})+\cos (\Delta \delta )
\label{eq:C}
\end{equation}
that we also expand in a power series: $C=1+C_{1}\lambda +C_{2}\lambda
^{2}+\ldots $

Finally, it follows from equations (\ref{eq:sin_phase}), (\ref{eq:y(x)}),
and (\ref{eq:C}) that
\begin{equation}
y_{j}(x)=C_{j}y_{0}(x)+y_{0}(x)\int_{1}^{x}\frac{dx^{\prime }}{%
y_{0}(x^{\prime })^{2}}\int_{0}^{x^{\prime }}\left( v_{1}y_{j-1}y_{0}\right)
(x^{\prime \prime })\,dx^{\prime \prime }  \label{eq:yj(x)}
\end{equation}
\begin{eqnarray}
\delta _{1} &=&-k\int_{0}^{1}v_{1}(x)y_{0}(x)^{2}\,dx  \nonumber \\
\delta _{2} &=&-k\int_{0}^{1}v_{1}(x)y_{0}(x)y_{1}(x)\,dx  \nonumber \\
\delta _{3} &=&\frac{\delta _{1}^{3}}{6}-k%
\int_{0}^{1}v_{1}(x)y_{0}(x)y_{2}(x)\,dx  \label{eq:delta_j}
\end{eqnarray}
and
\begin{eqnarray}
C_{1} &=&\delta _{1}\cot (k+\delta _{0})  \nonumber \\
C_{2} &=&\delta _{2}\cot (k+\delta _{0})-\frac{\delta _{1}^{2}}{2}  \nonumber
\\
C_{3} &=&\left( \delta _{3}-\frac{\delta _{1}^{3}}{6}\right) \cot (k+\delta
_{0})-\delta _{1}\delta _{2}.  \label{eq:Cj}
\end{eqnarray}
At the jth perturbation step we first solve for $y_{j}(x)$ and $\delta _{j}$
and then calculate $C_{j}$. Those expressions are sufficient for present
discussion, one can easily derive as many as necessary from the equations
above.

The number of necessary perturbation corrections depends on the convergence
rate of the perturbation series. For that reason it is important to choose a
convenient unperturbed or reference potential $v_{0}(x)$. It has been argued
that in order to have an adequate rate of convercence the unperturbed and
perturbed potentials should support the same number of bound states\cite
{ACCLY92}. This conclusion is based on Levinson's theorem than in our case
takes the form $\delta (k=0)=N\pi $, where $N$ is the number of bound states%
\cite{N66}. Here we investigate the application of the Bargmann--Schwinger
upper limit to the number of bound states \cite{B52,S61} and require that at
least
\begin{equation}
\int_{0}^{1}v_{0}(x)\,dx=\int_{0}^{1}v(x)\,dx.  \label{eq:BS_limit}
\end{equation}
Although this condition does not completely guarantee that both potentials
have the same number of bound states, it is a reasonable choice that will
prove sound in the example below.

The simplest exactly solvable unperturbed model is one with a constant
potential
\begin{equation}
v_{0}(x)=\left\{
\begin{array}{c}
-v_{0}\;if\;0<x<1 \\
0\;if\;x\geq 1
\end{array}
\right.  \label{eq:v0_const}
\end{equation}
in which case
\[
v_{0}=-\int_{0}^{1}v(x)\,dx.
\]
The starting--point of the calculation are the unperturbed solutions
\begin{eqnarray}
y_{0}(x) &=&\frac{\sin (k+\delta _{0})}{k\sin (K)}\sin (Kx),\;0<x<1,
\nonumber \\
K &=&\sqrt{k^{2}+v_{0}},  \nonumber \\
\delta _{0} &=&\arctan \left( \frac{k\tan (K)}{K}\right) -k+n\pi
\label{eq:unpert_sol}
\end{eqnarray}
where we choose $n$ so that $\delta _{0}(k\rightarrow \infty )=0$ as stated
above.

\section{Example \label{sec:example}}

In what follows we compare the perturbation results with accurate numerical
phase shifts calculated by a straightforward power--series method. To this
end we assume that the Taylor series of the potential around $x=0$, $%
v(x)=w_{0}+w_{1}x+\ldots $ converges at $x=1$, and expand the solution in
the same way: $y(k,x)=c_{0}(k)+c_{1}(k)x+c_{2}(k)x^{2}+\ldots $, where $%
c_{0}=c_{2}=0$ and $c_{j}(k,c_{1})=c_{j}(k,1)c_{1}$, $j=3,4,\ldots $.
Finally, we obtain the phase shift as
\begin{equation}
\delta =\arctan \left( \frac{y(k,1)}{y^{\prime }(k,1)}\right) -k+n\pi .
\label{eq:delta_example}
\end{equation}

As an illustrative example, we choose
\begin{equation}
v(x)=Ax(x-1),\;A>0,\;0<x<1  \label{eq:v_example}
\end{equation}
that can be solved in terms of confluent hypergeometric functions\cite{AS72}%
. However, here we resort to the power series approach because it is
suitable for more general problems and converges fast in most cases. The
power--series approach also enables us to obtain the bound--state energies
from the conditon
\begin{equation}
y^{\prime }(k,1)+ky(k,1)=0,\;k=\sqrt{-\epsilon }  \label{eq:BS_energy}
\end{equation}

We choose a well of depth $v_{0}$ as unperturbed model and the simple
condition (\ref{eq:BS_limit}) gives us $v_{0}=A/6$. We could obtain
analytical expressions for $\delta _{1}(A,v_{0},\delta _{0})$, $\delta
_{2}(A,v_{0},\delta _{0})$ and $\delta _{3}(A,v_{0},\delta _{0})$ but we
only show the first perturbation correction because the other ones are too
long.
\begin{eqnarray}
\delta _{1} &=&\frac{\sin (\delta _{0}+k)^{2}}{12k(k^{2}+v_{0})^{3/2}\sin
\left( \sqrt{k^{2}+v_{0}}\right) ^{2}}  \nonumber \\
&&\left\{ 3\left[ a-2v_{0}\left( k^{2}+v_{0}\right) \right] \sin \left(
\sqrt{k^{2}+v_{0}}\right) \cos \left( \sqrt{k^{2}+v_{0}}\right) \right.
\nonumber \\
&&\left. +\sqrt{k^{2}+v_{0}}\left[ 3a\sin \left( \sqrt{k^{2}+v_{0}}\right)
^{2}-a\left( k^{2}+v_{0}+3\right) \right. \right.  \nonumber \\
&&\left. \left. +6v_{0}\left( k^{2}+v_{0}\right) \right] \right\}
\label{eq:model_delta1}
\end{eqnarray}
In order to test the accuracy of perturbation theory we choose parabolic
wells with $A=6$ with no bound states, and $A=18$ with one bound state.
Figures 1 and 2 show the logarithmic error $\epsilon _{\log }=\log |(\delta
_{PS}-\delta _{PT})/\delta _{PS}|$, where $PS$ and $PT$ denote power series
and perturbation theory, respectively. We clearly appreciate that the
perturbation series converges for both parabolic--well strengths, and for
all energies. For some $k$ values the perturbation series of order $j-1$ may
be more accurate than the one of order $j$. We do not know if this behaviour
is fortuitous or if there is a mathematical or physical reason behind. Table~%
\ref{tab:err_av} shows the average relative error $\epsilon _{av}=\frac{1}{M}%
\sum_{j=1}^{M}|[\delta _{PS}(k_{j})-\delta _{PT}(k_{j})]/\delta
_{PS}(k_{j})| $ for some $A$--values. We clearly appreciate that the
perturbation series converges smoothly.

\section{More general problems \label{sec:general_problems}}

One can easily apply the method of Dalgarno and Steward to more general
problems of the form
\begin{equation}
y^{\prime \prime }(x)=Q(y,x)  \label{eq:y"_gen}
\end{equation}
where $Q(y,x)$ denotes a differential operation on $y(x)$. We choose a
convenient unperturbed problem
\begin{equation}
y_{0}^{\prime \prime }(x)=Q_{0}(y_{0},x)  \label{eq:y"_0_gen}
\end{equation}
and proceed as in Sec.~\ref{sec:model}. We obtain the following general
equation
\begin{eqnarray}
y(x) &=&C_{2}y_{0}(x)+C_{1}y_{0}(x)\int_{\beta }^{x}\frac{dx^{\prime }}{%
y_{0}(x^{\prime })^{2}}  \nonumber \\
&&+y_{0}(x)\int_{\beta }^{x}\frac{dx^{\prime }}{y_{0}(x^{\prime })^{2}}%
\int_{\alpha }^{x}\left( Qy_{0}-Q_{0}y\right) (x^{\prime \prime
})\,dx^{\prime \prime }  \label{eq:y(x)_gen}
\end{eqnarray}
where we can choose the arbitrary integration limits $\alpha $ and $\beta $
and constants $C_{1}$, and $C_{2}$, conveniently according to the problem.

Notice that if $Q_{0}(y_{0},x)=Q_{0}(x)y_{0}(x)$ then
\begin{equation}
u(x)=y_{0}(x)\int_{\beta }^{x}\frac{dx^{\prime }}{y_{0}(x^{\prime })^{2}}
\label{eq:u(x)}
\end{equation}
is a solution of $u^{\prime \prime }(x)=Q_{0}(x)u(x)$ and $W(u,y_{0})=1$.
That is to say, $y_{0}(x)$ and $u(x)$ are two independent solutions of the
unperturbed model.

It is not difficult to verify that the method of Dalgarno and Lewis applies
to velocity--dependent problems\cite{J06}. If we choose
\begin{eqnarray}
Q(y,x) &=&[v_{0}(x)+\lambda v_{1}(x)-\epsilon ]y(x)+\lambda \rho
(x)y^{\prime \prime }(x)  \nonumber \\
&&-\lambda \left[ y^{\prime }(x)-\frac{y(x)}{x}\right] \rho ^{\prime }(x)
\label{eq:Q_Jag}
\end{eqnarray}
then we have the model treated by Jaghoub\cite{J06}. Therefore, if
\begin{equation}
Q_{0}(y_{0},x)=[v_{0}(x)-\epsilon _{0}]y_{0}(x)  \label{eq:Q0_Jag}
\end{equation}
we have
\begin{eqnarray}
Qy_{0}-Q_{0}y &=&-\Delta \epsilon yy_{0}+\lambda v_{1}yy_{0}+\lambda \rho
y_{0}y^{\prime \prime }  \nonumber \\
&&-\lambda \left[ y^{\prime }-\frac{y}{x}\right] \rho ^{\prime }y_{0}.
\label{eq:(Qy0-Q0y)_Jag}
\end{eqnarray}
If we expand $y(x)$ as in Eq. (\ref{eq:PT_series}) and $\Delta \epsilon
=\epsilon _{1}\lambda +\epsilon _{2}\lambda ^{2}+\ldots $ then we obtain
Jaghoub's perturbation equations\cite{J06} provided that we choose the
arbitrary constants conveniently. In other words: Jaghoub's procedure is
merely the method of Dalgarno and Lewis for bound states developed in a
different way.

By means of the equations displayed above, the reader can easily verify that
present implementation of the method of Dalgarno and Lewis is also suitable
for the treatment of scattering problems with velocity--dependent
interactions\cite{RFL62}.

\begin{figure}
\begin{center}
\includegraphics[width=9cm]{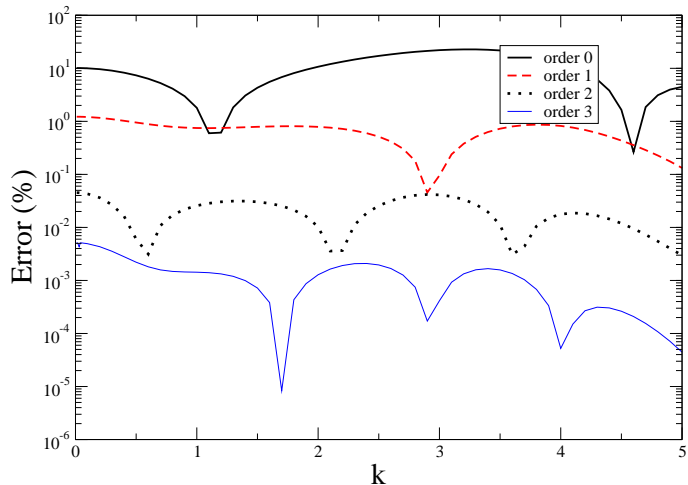}
\end{center}
\caption{The error $\epsilon = \left| \frac{\delta_{PS}-\delta_{PT}}{\delta_{PS}} \right| \times 100$ for $A=6$. }
\bigskip
\label{fig:A6}
\end{figure}

\begin{figure}
\begin{center}
\includegraphics[width=9cm]{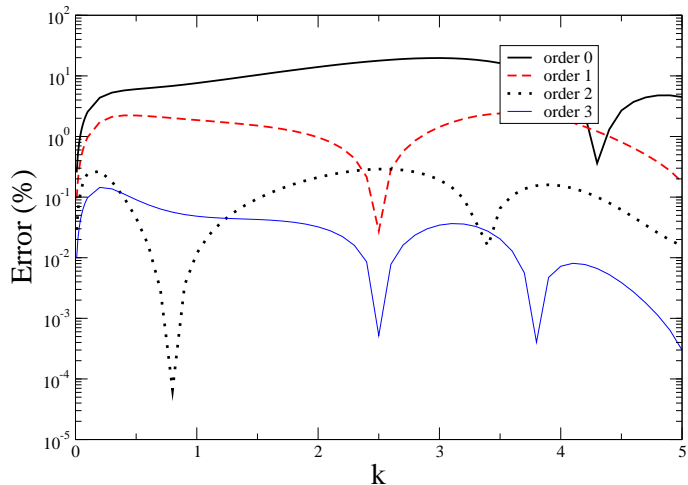}
\end{center}
\caption{The error $\epsilon = \left| \frac{\delta_{PS}-\delta_{PT}}{\delta_{PS}} \right| \times 100$ for $A=18$. }
\bigskip
\label{fig:A18}
\end{figure}

\section{Conclusions \label{sec:conclusions}}

We have shown that the method of Dalgarno and Lewis is suitable for the
calculation of phase shifts by means of perturbation theory. One can use
real eigenfunctions because their nodes do not cause the occurrence of poles
into the solutions.

The constant potential is a suitable unperturbed or reference model for the
application of perturbation theory to some problems. A simple prescription
for the well--depth leads to reasonable convergence rate and enables one to
obtain accurate results with perturbation series of low order.

We have shown that the method of Dalgarno and Lewis also applies to
bound--state and scattering problems with velocity--dependent interactions%
\cite{J06,RFL62}.

\begin{table}[tbp]
\caption{Average relative error for the parabolic well $V(x)=Ax(x-1)$}
\label{tab:err_av}
\begin{tabular}{rcccc}
$A$ & Zeroth order & First order & Second order & Third order \\
6 & 0.1060 & 0.00736 & 0.00024 & 0.000018 \\
12 & 0.0917 & 0.02250 & 0.00106 & 0.000672 \\
18 & 0.0906 & 0.01260 & 0.00127 & 0.000365 \\
24 & 0.0814 & 0.01109 & 0.00200 & 0.000327
\end{tabular}
\end{table}

\end{document}